\documentclass[10pt,journal,epsfig]{IEEEtran}
\usepackage[dvips]{graphicx}
\usepackage{graphicx}
\usepackage{amssymb}
\usepackage{cite,comment}
\usepackage{amsmath,color}
\usepackage{bm}
\makeatletter

\newcommand{\Rmnum}[1]{\expandafter\@slowromancap\romannumeral #1@}
\makeatother

\begin{document}
\IEEEoverridecommandlockouts
\title{Physical Layer Service Integration in 5G: Potentials and Challenges}
\author{Weidong Mei, Zhi Chen, \IEEEmembership{Senior Member, IEEE}, Jun Fang, \IEEEmembership{Member, IEEE}, and Shaoqian Li \IEEEmembership{Fellow, IEEE}
\thanks{This work was supported in part by the National Natural Science Foundation of China under Grant 61631004 and Grant 61571089.}
\thanks{W. Mei is with NUS Graduate School for Integrative Sciences and Engineering, National University of Singapore, Singapore 117456. He is also with the Department of Electrical and Computer Engineering, National University of Singapore, Singapore 117583 (e-mail: wmei@u.nus.edu).}
\thanks{Z. Chen, J. Fang, and S. Li are with National Key Laboratory of Science and Technology on Communications, University of Electronic Science and Technology of China, Chengdu 611731, China (e-mails: chenzhi@uestc.edu.cn; JunFang@uestc.edu.cn; lsq@uestc.edu.cn).}}
\maketitle

\begin{abstract}
High transmission rate and secure communication have been identified as the key targets that need to be effectively addressed by fifth generation (5G) wireless systems. In this context, the concept of physical-layer security becomes attractive, as it can establish perfect security using only the characteristics of wireless medium. Nonetheless, to further increase the spectral efficiency, an emerging concept, termed physical-layer service integration (PHY-SI), has been recognized as an effective means. Its basic idea is to combine multiple coexisting services, i.e., multicast/broadcast service and confidential service, into one integral service for one-time transmission at the transmitter side. This article first provides a tutorial on typical PHY-SI models. Furthermore, we propose some state-of-the-art solutions to improve the overall performance of PHY-SI in certain important communication scenarios. In particular, we highlight the extension of several concepts borrowed from conventional single-service communications, such as artificial noise (AN), eigenmode transmission etc., to the scenario of PHY-SI. These techniques are shown to be effective in the design of reliable and robust PHY-SI schemes. Finally, several potential research directions are identified for future work.
\end{abstract}

\section{Introduction}
In the last few years or so, streaming video services have played prominent roles in modern daily life. The global communications industry has been designing effective digital broadcast systems that cater to users' great demand for high spectral efficiency, driven by the massive deployment of mobile video services on modern large-screen devices. The goal of achieving high spectral efficiency has been considered as one of the key targets in fifth generation (5G) wireless systems. In fact, significant progress has been witnessed toward this goal since third generation (3G) wireless communication systems. In the 3rd Generation Partnership Project (3GPP) Release 9 standards, multimedia broadcast/multicast service (MBMS) has evolved to achieve improved performance with higher speed and more flexible service configuration, named as evolved MBMS (eMBMS)\cite{lecompte2012evolved}. In addition, the MBMS was also indispensable in Worldwide Interoperability for Microwave Acess (commonly known as WiMax)\cite{etemad2009multicast}.

Recently, the advent of multi-user multiple-input multiple-output (MU-MIMO) technique leads to a further increase in spectral efficiency. MU-MIMO combines the high capacity with the benefits of space-division multiple access, which is able to support 1000 $\times$ data rate in 5G \cite{andrew2014what}. MU-MIMO can overcome the bottleneck of receive degrees of freedom (d.o.f) in single-user MIMO. The capacity region for the MU-MIMO broadcast channel has been well analyzed using the dirty paper coding (DPC) techniques\cite{costa1983writing}. However, the main drawback of DPC lies in its prohibitively high operational complexity at both the transmitter and receiver. A more physically realizable solution to the multi-user transmission is based on linear processing, which assumes that the transmitted signal is generated by a linear combination of data symbols. An excellent overview of this solution is given in \cite{spencer2004an}. Besides transmitting private messages in MU-MIMO, the delivery of common message to multiple users, i.e., multicast service, is also widely needed when providing the services like video and audio. The multicast capacity for a general MU-MIMO channel can be obtained via convex optimization if ignoring the rank constraints. However, this results in high decoding complexity at each receiver. To seek an efficient transmit solution, many works have paid attention to the single-stream beamforming design in multiple-input single-output (MISO) channels; see, for example, \cite{sidi2006transmit,kim2011optimal,wu2013physical}.

On the other hand, the emerging application scenarios with mass data, such as e-banking, e-commerce, and medical information transfer, often pose various security threats to the traffic data. For instance, the service providers tend to offer fee-based services due to the copyright law. Apparently, the paid service data should be kept perfectly secure from the unauthorized users. Traditionally, the system designers often resorted to upper-layer protocols, typically, the cryptography technique to guarantee the overall security. Unfortunately, the large-scale use of the cryptography technique is unrealistic in e.g., mobile or ad-hoc networks due to the following reasons.
\begin{enumerate}
  \item Most cryptographic methods used today, like RSA, though rely on the difficulty of certain problems, offer no guarantee against clever attacks with new algorithms in the future.
  \item RSA and other public-key cryptographic methods can be broken by any eavesdropper intercepting transmission given sufficient time and computational power\cite{boneh2000why}.
  \item Cryptography requires a secure channel for exchange of private keys. While in mobile or ad-hoc networks, it is a difficult task to provide a reliably secure channel, which may lead to security hole in the distribution of the keys.
\end{enumerate}
In view of these limitations, there have been other efforts at seeking new secure methods of communication. The concept of physical-layer security emerges in this context. Wyner's pioneering work \cite{wyner1975wire} lays the foundation for the technique of physical-layer security. Intuitively speaking, the idea of physical-layer security is to establish an SNR advantage at the legitimate receiver relative to the SNR of potential eavesdroppers. Physical-layer security avoids the heavy dependence on complicated upper-layer-based encryption by only exploiting the physical characteristics of wireless channels, which is particularly desired in resource-limited application scenarios. In practice, physical-layer security can serve as a promising complement to the conventional cryptographic techniques. For more detailed introduction to physical-layer security, readers are referred to the existing overviews \cite{shiu2011physical,hong2013enhancing,mukherjee2014principles,liu2016physical,chen2017survey} and the references therein.

The remarkable progress in the development of MBMS and physical-layer security leaves us an interesting question: Is it possible to combine these two concepts together to simultaneously achieve high spectral efficiency and high-level security? Specifically, it is necessary to investigate an efficient integration of multiple services (i.e., broadcast/multicast service and confidential service) at the physical layer. One can envision that the physical-layer service integration (PHY-SI) could find a wide range of applications in the commercial and military areas. Many commercial applications, e.g., advertisement, digital television, and so on, are supposed to provide personalized service customization. As a consequence, confidential (fee-based) service and broadcast/multicast (free) service are collectively provided, and different user groups subscribe to one service according to their preferences. In battlefield scenarios, it is essential to propagate commands with different security levels to the frontline. The public information should be distributed to all soldiers, while the confidential information can only be accessed by specific soldiers.

Owing to the special characteristics of PHY-SI, several interesting questions arise. For example, how to allocate resources between confidential messages and public messages? How to tackle the mutual interference between the two types of messages? What is the proper decoding scheme at the receivers? The answers to these questions cannot be found in existing papers on physical-layer security, which necessitates a systematic study on PHY-SI. Before proceeding, hereon we specially highlight some remarkable differences between physical-layer security and PHY-SI, which are summarized as below.
\begin{enumerate}
  \item \emph{Differences in transmitted signal}. In physical-layer security, \emph{each} transmitted message is confidential and subject to strict secrecy constraints. Any unauthorized receiver has no access to the transmitted signal. However, PHY-SI generalizes physical-layer security by enabling simultaneous confidential and public information transfer. At the transmitter side, the transmitted signal is obtained by superposition of the confidential messages and the public messages. Note that there is no secrecy constraint imposed upon the public messages, and thus, even the unauthorized receivers have partial access to the transmitted signal.
  \item \emph{Differences in design objective}. In physical-layer security, there only exists a single secrecy rate, and the associated design objective is simply to maximize the secrecy rate. While in PHY-SI, each type of services would result in one specific transmit rate. All of the achievable rate tuples form a secrecy capacity region. As such, the primary design objective in PHY-SI is to seek a Pareto optimal transmit scheme, such that the boundary of the secrecy capacity region can be attained or approached.
  \item \emph{Differences in optimal transmit strategy}. Depending on the encoding order, the confidential message may play a role of interference to the public messages. The conventional strategy aiming for secrecy rate maximization is thus not optimal in this sense. Most of the widely adopted transmit strategies in physical-layer security, e.g., artificial noise-aided transmission, generalized eigenvalue decomposition (GSVD)-based precoding, etc., should be reformulated (not necessarily abandoned) to accommodate the superimposed public messages, which is not a trivial issue.
  \item \emph{Differences in coding}. Besides the aforementioned differences, PHY-SI is also distinct from physical-layer security in terms of encoding and decoding schemes. Generally, the superposition coding is exploited at the transmitter side, and the receivers perform the successive interference cancellation to obtain their intended messages. Based on different scenarios, the optimal codebook for the transmitted messages are also different.
\end{enumerate}

Actually, there have been some research activities that analyze the topic of PHY-SI, but most of them tackle this problem from an information-theoretic point of view. The fundamental aim of these works is to obtain capacity results or to characterize coding strategies that lead to certain rate regions\cite{Schaefer2014Physical}. Generally, the rate regions are given by a union over all possible transmit covariance matrices satisfying certain power constraints. To bridge the gap between the previous information-theoretic results and practical implementation, there is growing interest in analyzing PHY-SI from a signal processing point of view recently. In particular, optimal or complexity-efficient transmit strategies have to be characterized, so that the achieved performance could reach/approach the boundary of the secrecy capacity region. Such strategies are usually given by optimization problems, which generally turn out to be nonconvex.

This article first introduces some typical models of PHY-SI developed in the information-theoretic literature in Section \Rmnum{2}. We also mention how to derive the capacity-achieving transmit design using different approaches. Considering that the traditional approach used in information-theoretic literature is time-consuming and inflexible, we introduce an alternative time-efficient approach proposed in our previous papers. In Section \Rmnum{2}, based on a typical model of PHY-SI, we introduce some state-of-the-art approaches in the field of PHY-SI so far. These approaches aim to fulfill the requirements arising from different application scenarios. Some approaches are borrowed from the traditional single-service communications. Our main purpose is to show that by judiciously reformulating these approaches, they are still able to yield satisfactory performance in a PHY-SI framework, in spite of the existence of public messages. Based on this fact, some new insights into the transmit design in PHY-SI can be drawn. Finally, this article points out where further studies are needed.

\section{Existing PHY-SI Models}
In this section, we intend to introduce how the existing PHY-SI models allow the efficient integration of public service and confidential service. To highlight the key ideas and insights, we restrict the discussion to the simplest possible scenario in each model. Nonetheless, the general insights are also valuable and relevant for a more complicated model.

\subsection{Integration of Multicast and Confidential Unicasting}
\begin{figure}[hbtp]
\centering
\includegraphics[width=3in]{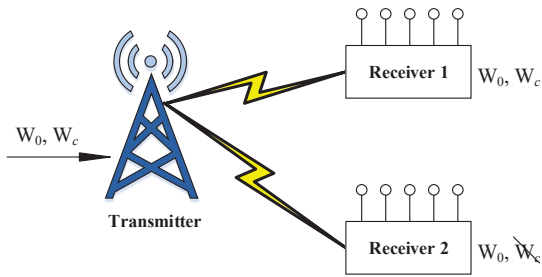}
\DeclareGraphicsExtensions.
\caption{PHY-SI of multicast and confidential unicasting}\label{Model1}
\end{figure}

This type of integration was first proposed in Csisz{\'a}r and K{\"o}rner's seminal work \cite{csiszar1978broadcast} under an information-theoretic name of Gaussian MIMO wiretap channel with common message. As shown in Fig.\,\ref{Model1}, in a basic model of this integration, a transmitter sends a common message $W_0$ to two receivers, and simultaneously, sends a confidential message $W_c$ intended only for receiver 1 and kept perfectly secure from receiver 2. We hereinafter name the intended (resp. unintended) receiver of the confidential message as an authorized (resp. unauthorized) receiver. Under discrete memoryless broadcast channel (DMBC) setup, Csisz{\'a}r and K{\"o}rner deduced the maximum rate region that can be applied reliably under the secrecy constraint (i.e., secrecy capacity region). Both the confidential message and the multicast message are encoded by using a standard Gaussian codebook. To achieve the secrecy capacity region, both receivers first decode their multicast message by treating the confidential message as noise. Then receiver 1 acquires a clean link for the transmission of its exclusive confidential message, without interference from the multicast message. In \cite{Hung2010Multiple}, the authors extended the results in \cite{csiszar1978broadcast} to the MIMO case by judiciously adopting a channel-enhancement argument.

This PHY-SI model is much more involved than the basic wiretap channel. In fact, it subsumes the basic wiretap model as a special case. This special case can be easily obtained from this PHY-SI model by disabling the multicast message. The optimal transmit design for the basic wiretap channel has been given in \cite{khisti2010secure1} and \cite{khisti2010secure2}, for MISO and MIMO case respectively. However, it seems to be more difficult to obtain the optimal transmit design for PHY-SI. The key problem of PHY-SI lies in how to integrate the two services as efficiently as possible to allow for large secrecy capacity region. Note that the two different service messages are coupled in the transmission. Along with this comes the unsatisfying fact that the existing approaches for secrecy rate maximization may not be applicable to PHY-SI. This is because the increase in the secrecy rate might lead to the performance degradation of the multicast service. Likewise, a strategy of pursuit in exclusive high multicast rate risks increasing the information leakage to the unauthorized receiver. It is thus of paramount importance to characterize the best tradeoff between these two services by appropriately designing the resource allocation and signal processing schemes.

\subsection{Integration of Multicast and Confidential Broadcasting}
\begin{figure}[hbtp]
\centering
\includegraphics[width=3in]{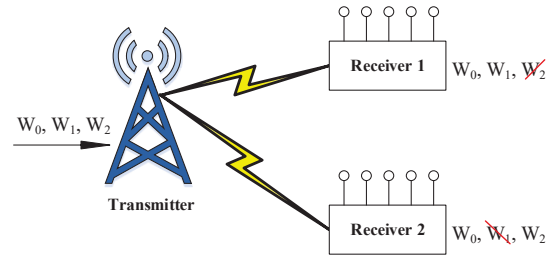}
\DeclareGraphicsExtensions.
\caption{PHY-SI of multicast and confidential broadcasting}\label{Model2}
\end{figure}

In the last subsection, it is assumed that only one user orders the confidential service. To fully exploit the potential of MU-MIMO, it is also possible to enable both receivers to order their own confidential service. To put it into context, in this subsection, we will introduce a more general integration scheme in which three messages $W_0$, $W_1$ and $W_2$ are delivered by the transmitter simultaneously, as shown in Fig.\,\ref{Model2}. Besides the multicast message $W_0$ intended for both users, $W_1$ ($W_2$) denotes the first (second) user's confidential message needing to be kept hidden from the second (first) user. We call the corresponding PHY-SI model as an integration of multicast and confidential broadcasting.

Note that the topic of confidential broadcasting itself has been studied in lots of literature. In information-theoretical works, confidential broadcasting is named as Gaussian MIMO broadcast channel with confidential messages. The secrecy capacity of confidential broadcasting was first established in \cite{liu2009secrecy}. The Gaussian codebook with the secrecy DPC (S-DPC) scheme was shown to be the optimal coding scheme \cite{liu2009secrecy}. In the special case of single receive antenna, the capacity-achieving transmit covariance matrices are both rank-one, which can be computed by the generalized eigenvalue decomposition. However, as for the general MIMO case, the capacity-achieving transmit covariance matrices can only be found by using an exhaustive search over a set of positive semidefinite matrices\cite{liu2009secrecy}, which is prohibitively complex. To circumvent the difficulty, several works propose some heuristic numeric optimization algorithms, say the block coordinate descent (BCD) algorithm in \cite{park2016weighted}, to obtain a lower bound on the secrecy capacity region. In addition, as an alternative to the optimal secret DPC scheme, a low-complexity linear precoding scheme was proposed in \cite{fakoorian2013on}. It was shown in \cite{fakoorian2013on} that linear precoding can be optimal if a specific condition is satisfied. All of the aforementioned literature restricts the number of users to two for analytical tractability. Actually, the secrecy capacity region for the general $K$-user case is still an open problem. The authors in \cite{geraci2012secrecy} deduced a conservative lower bound on the secrecy capacity region for this general case, and proposed a linear precoding scheme to maximize the sum secrecy rate.

The work \cite{ekrem2012capacity} and \cite{liu2013new} obtained the secrecy capacity region for this PHY-SI model almost at the same time. To achieve the secrecy capacity region, the multicast message should be encoded by using a standard Gaussian codebook, and the two confidential messages need to be encoded by using the S-DPC scheme developed in \cite{liu2009secrecy}. The two receivers still first decode the multicast message by treating the confidential messages as noise, and then each user decodes the confidential message intended to itself. Relying on the encoding order used in S-DPC, one of the users acquires a clean link for the transmission of its own confidential message, where there is no interference from the other user's confidential message. It is worth mentioning that the secrecy capacity region is proved to be invariant with respect to the encoding order used in S-DPC \cite{ekrem2012capacity}. To address the issue of transmit design arising from this PHY-SI model, we are still confronted with the problem of how to efficiently superimpose the messages with proper power levels and precoding schemes. With the multicast message, the generalized eigenvalue decomposition may not be optimal for the confidential information transfer in this model.

\subsection{PHY-SI in Bidirectional Relay Networks}\label{twoway}
In secure wireless networks, relay cooperation has been shown to be capable of extending network coverage and providing spatial diversity and secrecy gain, and thus has drawn much attention in the literature. Depending on the role of the relays, there are generally three kinds of relay-assisted designs in traditional physical-layer security, i.e., cooperative relaying, cooperative jamming, and hybrid relaying and jamming\cite{rodriguez2015physical,chen2015multi}. In the first case, the relays forward the information from the source to the legitimate receiver to strengthen the quality of the legitimate link. In cooperative jamming, relays cooperatively emit jamming signal to jam the eavesdroppers. By generalizing the above two designs, some papers put forth the hybrid relaying and jamming scheme, in which the relays are able to simultaneously forward information and emit jamming signal. Additionally, attention has also been paid to untrusted relay networks, in which the source-destination pair also wishes to keep the information confidential from the relays. Pioneering studies \cite{he2010cooperation,jeong2012joint} have shown that the security performance can still be improved by cooperating with untrusted nodes. Among various relay-aided models, bidirectional (two-way) relaying is becoming increasingly attractive recently due to its potential in further advancement of performance and coverage in wireless networks. By leveraging the property of bidirectional communication, bidirectional relaying is able to compensate the inherent loss in spectral efficiency induced by half-duplex relays\cite{liu2015band}.
\begin{figure}[hbtp]
\centering
\includegraphics[width=3in]{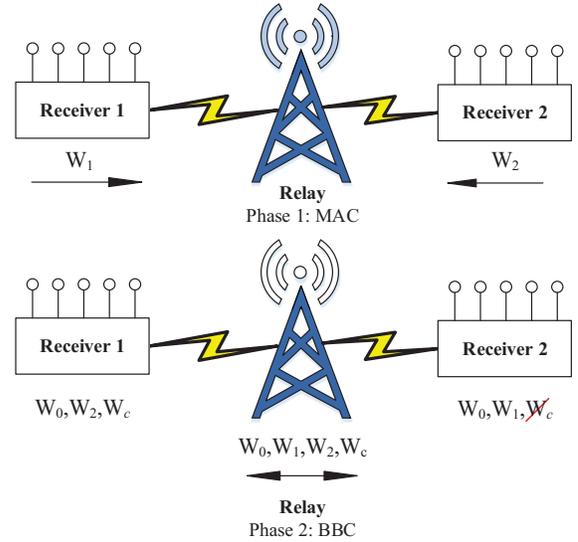}
\DeclareGraphicsExtensions.
\caption{PHY-SI in bidirectional relay networks}\label{Model3}
\end{figure}

The previous work \cite{Wyrembelski2012Physical} pointed out that it is feasible to implement PHY-SI in a bidirectional relay network. To show this, let us take a three-node bidirectional relay network as an example, where a half-duplex relay node establishes a bidirectional communication between two other nodes using a decode-and-forward protocol, as shown in Fig.\,\ref{Model3}. The relay transmissions are divided into two phases. In the initial multiple access (MAC) phase, two nodes send their messages $W_1$ and $W_2$ to the relay node and then the relay node decodes them. In the succeeding bidirectional broadcast (BBC) phase, the relay re-encodes and forwards $W_1$ and $W_2$ to the two receiving nodes. To realize PHY-SI, the relay further integrates a common message $W_0$ for both receivers and a confidential message $W_c$ for receiver 1 only, of which receiver 2 should be kept ignorant. As seen, this PHY-SI model manages to integrate private, common, and confidential messages at PHY.

Here we should specially mention some properties of this PHY-SI model. First of all, it is possible for the receiving nodes to decode the other's message in the BBC phase using their own message from the MAC phase as side information. Specifically, for each node, the information transmitted by itself in the first phase performs as self-interference during the second phase. Since each node has partial knowledge about the transmitted signal from the relay, it is able to subtract its own message from the received signal to cancel the self-interference. After that, it can detect the information from the other node. To distinguish this process from the traditional broadcasting, the second phase is named as BBC phase by literature. Moreover, by following the network coding idea, the secrecy capacity region can be achieved by only transmitting a single data stream that combines both private messages at the relay side. In addition, the side information at the receivers allows us to use a rate-splitting approach between the common rate and the individual rates. Consequently, only two transmit covariance matrices, i.e., the covariance matrices of the confidential message and the public messages, are involved in the secrecy capacity region in this case.

\subsection{Computation of Optimal Transmit Covariances}\label{comp}
It is always desirable to obtain a transmission scheme whose performance attains the boundary of the secrecy capacity regions, since, otherwise, the achievable rates of some of the services can be further improved. Hence, to determine the optimal transmission scheme, it is required to seek the boundary points of the secrecy capacity regions. To obtain the associated transmit covariance matrices, an optimization problem, termed secrecy capacity region maximization (SCRM) problem, has to be solved, which contains multiple objectives and is generally nonconvex.

There exist several typical approaches to finding the boundary points of a secrecy capacity region. The first approach is referred to as re-parameterization\cite{Schaefer2014Physical}, which is extensively exploited in the information-theoretic works. Its basic idea is to introduce an auxiliary variable and a weight vector to reexpress the achievable rate of each message. This approach is able to yield all boundary points by varying the weights and solving a convex feasibility problem. Even so, the computational complexity of such an approach is rather high since solving a feasibility problem requires a bisection search. Besides, this re-parameterizing approach is only applicable to a simple two-user MISO case; it would become intractable or prohibitively time-consuming to use this approach if the number of users and/or receive antenna increases.

To address the limitation, an alternative approach was proposed recently in \cite{mei2016secrecy,mei2017on}. Specifically, this approach directly attacks the SCRM problem through the convex optimization techniques. As is known, a standard technique for dealing with the multi-objective optimization problem is scalarization, by which the Pareto optimal solutions can be found. One feasible method of scalarization is to move some of the objectives to the set of constraints and constrain each of them by some constants. Then, by varying the constants, tradeoffs between different objectives can be struck. In our previous works\cite{mei2016secrecy,mei2017on}, it has been shown that this method can significantly reduce the computational complexity compared to the re-parameterizing method. Another possible method of scalarization is to maximize the weighted sum of different objectives. By varying the weight vector and maximizing the corresponding weighted sum, one can acquire different Pareto optimal points. However, this method relies critically on the convexity of the rate region. Unless the rate region is strictly convex, this method can only yield a portion of all Pareto optimal points.

\section{State-of-the-Art Techniques for PHY-SI}
PHY-SI faces a variety of challenging issues, the centerpiece of which is how to guarantee the quality of the confidential service while not compromising that of the public services. So, it is possible and necessary to adopt some physical-layer security techniques to enhance the overall performance of PHY-SI. Nonetheless, different from traditional physical-layer security techniques, the techniques designed for PHY-SI concentrate not only on the confidential service, but also the other services. We hereinafter introduce several state-of-the-art techniques for efficient realization of PHY-SI. Unless otherwise specified, the integration of multicast and confidential unicasting will be used as a basic model in this section.

\subsection{Artificial Noise}\label{ArtflNs}
Goel and Negi first coined the term artificial noise (AN)\cite{goel2008guaranteeing}. In AN-aided transmission, the transmitter sends artificially generated noise to interfere the eavesdroppers deliberately. This active scheme has been demonstrated as an effective means to safeguard the overall network in multi-antenna systems. In previous papers, it has been revealed that the way of generating AN in fact depends critically on several factors detailed as below.
\begin{enumerate}
  \item First, the number of eavesdroppers. For the special case with a single eavesdropper, transmit design without AN has sufficed to achieve the secrecy capacity\cite{khisti2010secure1,khisti2010secure2}. However, in the general case with multiple eavesdroppers, the AN-aided transmit design is capable of offering significant secrecy gain over the no-AN one\cite{li2013spatially,li2013transmit}. This is because AN can compensate the transmit d.o.f. bottleneck at the transmitter.
  \item Second, the availability of eavesdroppers' CSI. As suggested by Goel and Negi, if only the statistics of eavesdroppers' CSI are known to the transmitter, a relative SNR advantage could still be realized by uniformly spreading AN in a subspace orthogonal to the main channel. To move a step further, if the eavesdroppers' CSI is partially known, additional gains may be achieved by optimizing the AN transmit covariance\cite{li2013spatially}.
  \item Third, the availability of eavesdroppers' locations. Though this factor could be regarded as a special case of the last factor, it breeds a great number of effort towards physical-layer security from the perspective of stochastic geometry. In \cite{zheng2015multi}, the authors showed that more power should be allocated to AN to achieve a higher secrecy rate in dense eavesdropper scenarios, given that the eavesdroppers' locations follow Poisson point process (PPP).
\end{enumerate}

In PHY-SI, the AN technique can be implemented much more effectively to cripple the reception of the unauthorized receivers. This is because the unauthorized receivers are all network users, which makes it possible to get their perfect CSI using, for example, pilot-symbol aided modulation. With perfect CSI on all links, there is a possibility for the transmitter to generate spatially selective AN to further enhance the performance of PHY-SI by optimizing the AN transmit covariance. Nonetheless, we will show shortly that the way of generating AN relies on a new factor, i.e., the multicast rate in PHY-SI. As mentioned earlier, in PHY-SI, the quality of public service should be further considered. It is evident to see AN exerts a negative effect on the performance of public services. Accordingly, AN is a double-edged sword in PHY-SI: Excessive use of AN will degrade the quality of public service at all receivers, while limited use of it cannot attain the best security performance. Consequently, the key in designing AN in PHY-SI is to adjust its transmit direction in order to achieve a dual balance, i.e., the balance between the interference to unauthorized receivers and that to the authorized receiver, and the balance between the performance of confidential service and that of public services.

\begin{figure}[hbtp]
\centering
\includegraphics[width=3.4in]{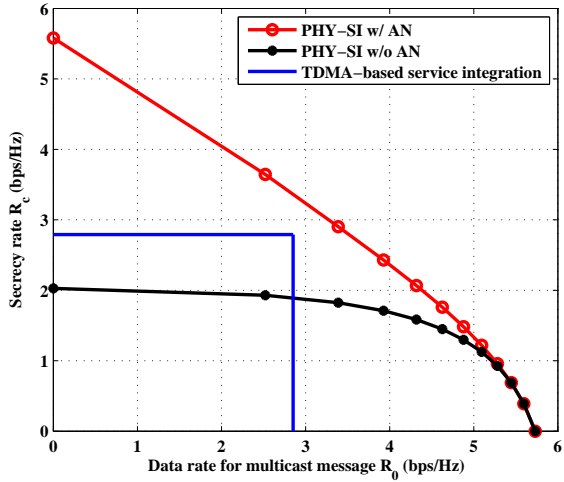}
\DeclareGraphicsExtensions.
\caption{Secrecy rate regions with and without AN\cite{mei2017on}}\label{SRR_pf}
\end{figure}
Fig. \ref{SRR_pf} depicts a simple example of MISO PHY-SI system in the presence of multiple unauthorized receivers. The number of transmit antenna is $N_t=2$, the number of unauthorized receivers is $K=4$, and the transmit SNR is 20dB. We show the achievable secrecy rate regions with and without AN in Fig. \ref{SRR_pf}. As seen, secrecy rates with AN are mostly higher than those without AN. The striking gap indicates that AN indeed enhances the security performance without compromising the quality of multicast service (QoMS). This is consistent with the previous observations in traditional physical-layer security. In fact, even the TDMA-based service integration\footnote{The TDMA-based service integration decouples the confidential service and multicast service by assigning them to two orthogonal time slots. Each type of service exclusively occupies the given time slot to distribute its message.} can provide better secrecy rate performance than the PHY-SI without AN at low QoMS region. This phenomenon can be interpreted from the transmit d.o.f.. The total d.o.f. of the unauthorized receivers is $K=4$, higher than the transmit d.o.f. $N_t=2$. The lack of transmit d.o.f. to deal with the unauthorized receivers is the reason for the unsatisfactory security performance of PHY-SI without AN. Nonetheless, with the increasing demand for QoMS, the gap between the AN-aided PHY-SI and the no-AN PHY-SI is gradually shortened to zero, which implies that AN is prohibitive at high QoMS region. This phenomenon never occurs in traditional physical-layer security and reveals a new insight that the use of AN is strictly subject to the demand for QoMS in PHY-SI. In addition, as expected, the AN-aided PHY-SI yields a significantly larger secrecy rate region than the TDMA-based service integration, which verifies that PHY-SI is able to offer ``integration gain''.

\subsection{Decoupling of Service Messages}
Due to the coupling of service messages leading to high nonconvexity of optimization problems, it becomes intractable to seek capacity-achieving transmit covariance matrices in MIMO systems. It is a natural idea to decouple the service messages in some domain if possible. This subsection will introduce an efficient ``eigenmode'' transmission scheme, i.e., GSVD-based precoding, which decouples the confidential message and multicast message in the spatial domain. In eigenmode transmission, the transceiver design is based on the SVD/GSVD of the MIMO channels. Through the decomposition, the MIMO channels can be decoupled into several parallel single-antenna channels, and the linear precoding design is reduced to the power allocation over parallel channels. In \cite{telatar1999capacity}, Telatar has shown that the eigenmode transmission is capacity-achieving in single-user MIMO systems, and that the optimal power allocation is spatial water-filling. Furthermore, in \cite{khisti2010secure2} and \cite{fakoorian2012optimal}, the authors studied the performance of eigenmode transmission in MIMO wiretap channels. Interestingly, based on the GSVD of the main and wiretap channels, the resulting transmission scheme was shown to be optimal in the high SNR regime. But the optimal power allocation is not in a water-filling form.

In GSVD-based MIMO PHY-SI, the communication links are decomposed into parallel eigenchannels that can convey \emph{either} confidential message \emph{or} multicast message. At the output of each eigenchannel, it is easy for the receivers to extract their desired signals by receive beamforming, as the eigenchannels are perfectly noninterfering. Notice that this GSVD-based PHY-SI scheme is a natural extension of the conventional eigenmode transmission proposed in \cite{telatar1999capacity,khisti2010secure2} and \cite{fakoorian2012optimal} to the case with multiple sorts of messages. Nonetheless, the resulting power allocation is significantly different in nature compared with that in \cite{telatar1999capacity,khisti2010secure2} and \cite{fakoorian2012optimal}. Specifically, due to the constraint on QoMS, the optimal power allocation has no closed-form solution. More importantly, in addition to the power allocation over parallel eigenchannels, a novel eigenchannel-service association problem occurs---that is, it is also required to associate each service message with an appropriate eigenchannel\cite{mei2016GSVD}. To make the problem more complicated, different demands on multicasting performance may correspond to different eigenchannel-service association, making this association dynamic. The resulting problem, finally, turns out to be a mixed-integer nonlinear optimization problem.

\begin{figure}[hbtp]
\centering
\includegraphics[width=3.4in]{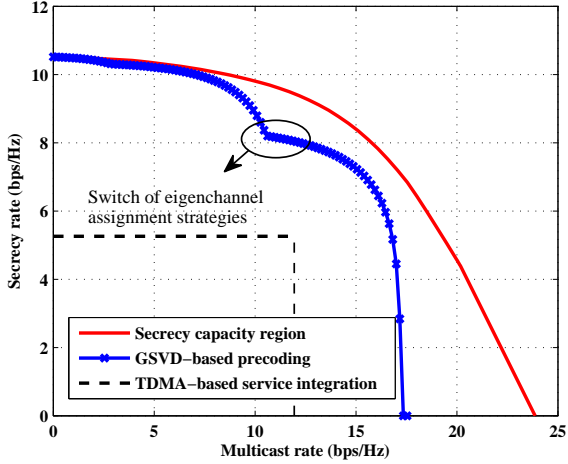}
\DeclareGraphicsExtensions.
\caption{Secrecy rate regions by message decoupling\cite{mei2016GSVD}}\label{decouple}
\end{figure}
Fig.\,\ref{decouple} plots the secrecy rate regions achieved by the GSVD-based PHY-SI scheme. The number of antennas at the transmitter, authorized receiver and unauthorized receiver are $N_t=5$, $N_b=4$ and $N_e=4$, respectively. The transmit SNR is 20dB. The secrecy capacity region is obtained from exhaustive search, and serves as a reference indicating the performance loss the GSVD-based scheme would inevitably experience. One can notice that there exists a switching point at the boundary of the GSVD secrecy rate region, which signals the switch of the eigenchannel assignment strategy. This switch results from the increasing demand for QoMS. If the current eigenchannel assignment strategy cannot sustain the demand for QoMS, the switch will occur, and the current eigenchannel assignment strategy will be switched to another one. One can also find that at low QoMS region, the GSVD-based scheme has identical performance to the exhaustive search. This is attributed to the near-optimality of GSVD-based precoding at high SNR in the confidential message transmission\cite{khisti2010secure2}. However, with the increase in QoMS, the gap between these two regions gradually expands. This performance degradation is due to the suboptimality of GSVD-based precoding to the multicast message transmission. Finally, it can be observed that the GSVD-based scheme suffices to give much better performance than the TDMA-based one.

\subsection{Robust PHY-SI}
In traditional physical-layer security, many reasons lead to imperfection of CSI at communication nodes, such as the following ones\cite{he2013wireless}:
\begin{enumerate}
  \item \emph{No feedback from the eavesdroppers}. If the eavesdropper is a passive entity, it would be difficult to secure its CSI or even location information. A widely adopted assumption for this case is such that the transmitter is aware of the distribution, first moment, and second moment of the eavesdropper's CSI.
  \item \emph{Imperfect feedback links}. If the feedback links are not error-free, noise component may be added to the feedback information. On the other hand, if the feedback links have a long delay, the CSI at the transmitter may be outdated.
  \item \emph{Channel estimation errors at the receivers}. Since the estimation of fading channels is in general not errorless, the CSI feedback from the receivers is therefore imperfect. Besides, as the quantized CSI is fed back from the receivers, the quantization error also has an impact on the average rate.
\end{enumerate}
Notice that PHY-SI guarantees that an unauthorized receiver will report its CSI to the transmitter for acquiring the public service, the issue in the first case can be resolved as a result. However, the acquisition of perfect CSI is still subject to the issues given in the last two cases. Thus, it is still a nontrivial problem to design PHY-SI schemes with robustness.

A viable approach to modeling the channel uncertainty in the literature is to bound the CSI estimation error in a spherical set and optimize the worst-case performance of the scheme. This approach isolates specific channel estimation methods from the resource allocation algorithm design by adjusting the size of the spherical set. Under this setup, the secrecy rate is limited by the worst channel to the authorized receiver and the best channel to the unauthorized receivers, while the multicast rate is limited by the worst channel to all receivers. The resulting secrecy rate region maximization problem is semi-infinite by nature due to the presence of the continuous channel uncertainties. The use of AN with imperfect CSI is also worth studying. It has been revealed that incorporating AN can effectively combat the performance degradation incurred by CSI imperfection in physical-layer security\cite{li2013spatially}. At this point, a critical question yet remains: is AN still beneficial in PHY-SI in consideration of the channel estimation error.

\begin{figure}[hbtp]
\centering
\includegraphics[width=3.4in]{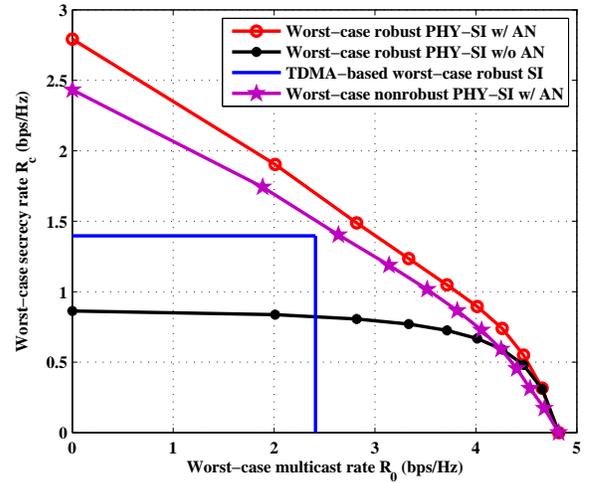}
\DeclareGraphicsExtensions.
\caption{Worst-case secrecy rate regions with and without AN\cite{mei2017on}}\label{robust}
\end{figure}
Fig.\,\ref{robust} depicts the worst-case secrecy rate regions with and without AN. All simulation parameters are consistent with those in Fig.\,\ref{SRR_pf} except the CSI uncertainty level. To show the superiority of the robust design, Fig.\,\ref{robust} also gives the worst-case secrecy rate region achieved by a nonrobust design, which utilizes the estimated CSI to optimize the precoder, regardless of the estimation error. Compared to Fig.\,\ref{SRR_pf}, we can clearly observe that the existence of channel uncertainty dramatically diminishes the achievable secrecy rate regions. One can also note the coincidence of the AN-aided scheme and the no-AN scheme at high QoMS region. This reveals that the use of AN still depends heavily on QoMS even though CSI uncertainty is considered. Finally, it should be mentioned that the robust PHY-SI scheme outperforms its nonrobust counterpart over the whole range of multicast rate.

Besides the worst-case robust approach, there exist some other design approaches for modeling the CSI error in the literature. The second approach assumes a probabilistic CSI error model and aims at good on-average performance, as opposed to the good worst-case performance. The third is the outage-based approach, whose design focus is on constraining QoS outages still under a probabilistic CSI error model. In contrast to the on-average performance, this approach seeks to provide ``safe'' performance, ensuring a certain chance of success of QoS deliveries. The outage-based approach is essential in delay-sensitive or low-latency applications. These two probability-based approaches often amounts to solving stochastic optimization problems. It is also interesting to study the performance of PHY-SI under these two CSI error models in the future.

\subsection{Energy-Efficient PHY-SI}
The requirement of dual or multiple high-rate transmission in PHY-SI significantly intensifies the need for energy compared to the transmission of a single message. The performance of PHY-SI is fundamentally constrained by the limited energy. To address this issue, energy-efficient PHY-SI scheme is a promising solution. In fact, energy efficiency (EE), measured in bits-per-Joule, has emerged as a new prominent figure of merit and has become one of the most widely adopted design metrics for realizing green communication in 5G nowadays. Extensive works have paid attention to the fundamental spectral-energy efficiency tradeoff, and readers are referred to the survey papers\cite{wu2014green,wu2017overview,zhang2017fundamental} and the references therein for more details. Notable results include the utilization of d.o.f in different resource domains, say time, space, power, bandwidth, and user, to balance EE and spectral efficiency. Among the various EE-enhanced solutions, here we specially mention the multi-antenna technique. By leveraging the additional spatial d.o.f in multiple-antenna system, the transmission rate can be significantly increased without additional bandwidth. The optimal transmit covariance matrices to achieve the fundamental limit of EE under static/fast fading/slow fading MIMO channel condition have been derived in \cite{belmega2011energy}. It was shown in \cite{belmega2011energy} that in static and fast fading channels, the EE is maximized at ultra-low transmit power. In practical systems, the circuit power consumption should also be included for energy efficiency maximization. The authors in \cite{xu2013energy} proposed a block-coordinate ascent algorithm to find the fundamental limit of EE with circuit power consumption under static MIMO channel condition. The proposed algorithm admits a water-filling-based solution in each iteration, and is guaranteed to converge to a globally optimal solution.

Recently, many works have extended the concept of EE to physical-layer security and put forth the concept of secrecy EE, defined as the ratio of the achievable secrecy rate to the total power consumption. The fundamental limit of SEE in static and fast fading multi-antenna wiretap channels was studied in \cite{zhang2014energy} and \cite{zappone2016energy}, respectively. Due to the intractability of the target problems, both papers exploited a combination of fractional programming theory and sequential optimization theory to solve the problems. This approach gives an approximate solution with polynomial complexity. The authors in \cite{zappone2016energy} provides further insights into the use of AN for secrecy EE maximization. Since using AN will incur increased energy consumption due to digital signal processing operations, the authors revealed that the use of AN does not always improve the secrecy EE, which is in contrast to secrecy rate optimization, as introduced in Section \ref{ArtflNs}. Another branch of research activities focussed on the issue of outage secrecy EE maximization\cite{ng2012energy,chen2013energy}. By outage secrecy EE we mean the ratio of the outage secrecy capacity to the total consumed power. The outage-based secrecy EE is considered when the channels experience slow fading or when the transmitter can not obtain a reliable estimation of CSI.

While the topic of energy-efficient communications has been extensively studied, its systematic investigation for PHY-SI is still in its infancy. The energy-efficient transmission in PHY-SI takes on several new characteristics and can be broadly classified into two categories. The first one is still the conventional spectral-energy efficiency tradeoff. The associated scheme aims to satisfy the communication requirement with the minimum energy expenditure on communication-related functions, e.g., communication circuits, signal transmission, etc. To this end, the problem can be formulated as a secrecy EE maximization problem with a multicast rate constraint. The other category of energy-efficient operation aims to characterize the fundamental EE tradeoffs between different services. This service-oriented EE tradeoff is somewhat novel and particularly beneficial in e.g., wireless sensor networks, where sensor nodes may be required to propagate different sorts of service message when concurrent events happen. It is evident that an energy-efficient resource allocation among these services would be helpful in prolonging the battery life for the sensor nodes. To best describe this novel service-oriented EE tradeoff, one first needs to establish the associated EE region, and then, maximizes this region through multi-objective optimization techniques\cite{mei2017energy}, as mentioned in Section \ref{comp}. The fundamental tools to solving the secrecy EE maximization problems in PHY-SI are still the combination of fractional programming theory and sequential optimization theory.

\begin{figure}[!t]
\centering
\includegraphics[width=3.4in]{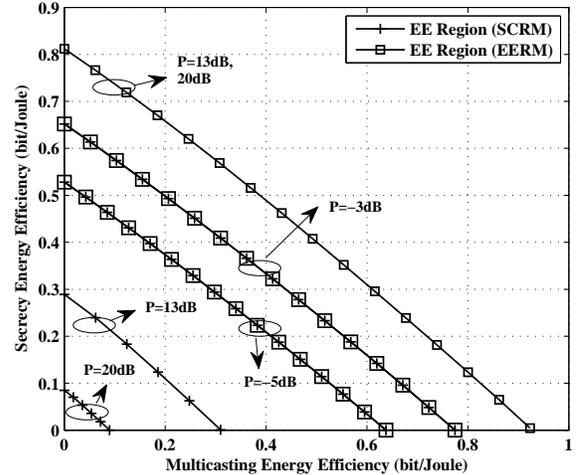}
\DeclareGraphicsExtensions.
\caption{EE regions in PHY-SI\cite{mei2017energy}}\label{EE_Region}
\end{figure}
Fig.\,\ref{EE_Region} illustrates the two-user EE regions achieved by different methods with different SNRs. Two strategies are considered and compared, i.e., SCRM strategy and EE region maximization (EERM) strategy. Fig.\,\ref{EE_Region} shows that the EE region achieved by EERM expands to its outermost boundary and eventually saturates for increasing transmit power. The reason for this phenomenon is that for high transmit power, the EERM strategy is not to transmit at full power. Consequently, when the transmit power grows large, the actual consumed power, however, does not grow large, but remains constantly equal to the unique and finite maximizer of the EE region. On the other hand, it can be observed that the EE region achieved by SCRM is also increasing and identical to that achieved by EERM in the beginning. However, when the transmit power is large, the former significantly diminishes and becomes inferior to the latter. This is because the SCRM strategy, unlike the EERM one, has to use up all available power to expand the secrecy capacity region, which undermines the EE performance accordingly in the high SNR regime. Finally, it is worth mentioning that the observations above agree with the previous results in \cite{zhang2014energy} and \cite{zappone2016energy}. Thus, the two-dimensional EERM problem is in essence a generalization of the traditional one-dimensional secrecy EE problem tackled by, for example, \cite{zhang2014energy} and \cite{zappone2016energy}.

\section{Future Directions}
PHY-SI will be a critical research topic in 5G, and there are a variety of challenging issues along which the developed results in this article can be further investigated. In what follows, we list some future research directions on PHY-SI. Here we remark that our focus in this section is on the application of the following directions to PHY-SI, even the directions themselves are not brand-new in wireless communications. The marriage of PHY-SI and the following timely directions may still provide some interesting and novel insights.

\subsection{PHY-SI for mmWave Large-Scale Antenna Arrays}
Our previous discussion on PHY-SI implicitly assumes that the transmitter adopts a digital beamforming method to realize PHY-SI. However, the fully digital beamforming methods require one radio frequency (RF) chain per antenna element, which makes it not economical for large-scale antenna arrays due to the high cost and high power consumption of RF chain components. To address this hardware limitation, existing papers propose the concept of analog beamforming and of hybrid analog and digital beamforming. The analog beamforming can be easily implemented with a single RF chain by using cheap phase shifters. In the hybrid beamforming, the beamformer is constructed by concatenation of a low-dimensional digital beamformer and an analog beamformer. It should be mentioned that even in the traditional physical-layer security, the analog or hybrid beamforming design for secrecy rate maximization is still an ongoing topic. In a very recent paper\cite{li2017constant}, the authors concentrated on the analog beamforming design to maximize the secrecy rate. The authors proposed a low-complexity numerical optimization approach to obtain an approximate solution of the analog beamformer. It was shown in \cite{li2017constant} that the obtained analog beamformer is capable of approaching the performance of a fully-digital scheme with much lower hardware complexity. Nonetheless, this result was obtained without integrating other types of services. If an additional multicast message is superimposed upon the confidential message, as highlighted before, the confidential message would interfere with the multicast message, and the QoMS constraint would further complicate the target optimization problem. Since the beamformer of the confidential message is involved not only in the term of secrecy rate, but also in the term of multicast rate, the approach proposed in \cite{li2017constant} cannot be directly used for PHY-SI with analog/hybrid beamforming. Notice that we have not taken into account several other components yet, such as AN, and the finite resolution of phase shifters. Thus, it is still an unsolved issue to find an effective and efficient signal-processing approach to enhance the performance of PHY-SI in mmWave communications.

\subsection{PHY-SI in OFDM/OFDMA System}
The previous works on PHY-SI are restricted to the flat-fading channels. However, PHY-SI is also expected to implement on broadband channels with frequency selectivity. Orthogonal frequency-division multiplexing (OFDM) is known to be robust against frequency-selective fading. In OFDM, a large number of closely spaced orthogonal sub-carrier signals are used to carry data on several parallel subchannels. As such, the crux of OFDM-based PHY-SI lies in how to appropriately allocate the transmission power, rate, and beamforming weight on a per-subcarrier basis. The achievable secrecy and multicast rates are both characterized by the sum of the achievable rate at each subchannel, which makes the SCRM problem more difficult to tackle. More interestingly, the use of OFDM provides a novel method to inject AN. To be specific, the AN can be added in the frequency domain and then transformed to the time domain with the service messages by inverse fast Fourier transform (IFFT). Alternatively, one can directly add AN in the time domain. The basic merit of time-domain AN lies in that even in a single-antenna network, AN may still boost the secrecy rate region by exploiting the temporal d.o.f. offered by the cyclic prefix in an OFDM symbol\cite{qin2013power}. Also note that as the time-domain AN will still interfere with the multicast message, it can only strengthen the secrecy rate in the low QoMS region.

Furthermore, orthogonal frequency-division multiple access (OFDMA) can be exploited to realize the integration of multicasting and confidential broadcasting, by allowing dynamic assignment of subcarriers to different receivers. In fact, a recent work \cite{benfarah2016power} has considered a similar setting and deduced an achievable secrecy rate region. However, the work \cite{benfarah2016power} failed to consider some practical aspects in OFDMA. The authors merely focussed on the power allocation problem under an assumption that all receivers have an access to each subcarrier for analytical simplicity. This general assumption also causes high operational complexity at the transmitter, as all confidential messages and a single multicast message should be superimposed at the input of each subchannel. Actually, in practice, one subcarrier is always allocated to at most one receiver, which imposes a nontrivial constraint on the subcarrier assignment strategy. By imposing such a constraint, the operational complexity at the transmitter can also be reduced, since at most two messages, including one confidential message intended for a specific receiver and one multicast message, need to be combined at the input of each subchannel. However, there is no free lunch. The joint power and subcarrier assignment optimization problem turns out to be a mixed-integer nonlinear problem and become even more intractable to handle in a PHY-SI framework. Finally, as the authors in \cite{benfarah2016power} only considered a single-antenna system setup, it is also interesting to exploit the spatial d.o.f by equipping the transceivers with multiple antennas.

\subsection{PHY-SI in Unmanned Aerial Vehicle (UAV) Communications}
UAV communications are expected to play a crucial role in the forthcoming 5G wireless networks, thanks to its many advantages such as low cost, high mobility, and on-demand deployment. The mobility of UAV and the line-of-sight (LoS) nature of air-to-ground channels also provide novel d.o.f. to facilitate the implementation of PHY-SI. Assume that a UAV plays a role of transmitter. To enhance the secrecy rate, the high mobility of UAV can be exploited to proactively establish stronger links with the authorized ground nodes and/or degrade the channels of the unauthorized nodes, by adjusting the distance between the UAV and the ground nodes, via proper trajectory design. The mobility of UAV can also be exploited to overcome the bottleneck-user issue arising from physical-layer multicasting. As is known, in a multicast channel, the multicast capacity is fundamentally limited by the user with the worst channel condition. If one user is located much farther away from the transmitter than the others, the multicast capacity will be fundamentally constrained by this user. The UAV can resolve this issue by adjusting its location over time to improve the wireless channels to different users at different locations. As such, UAV should be a promising candidate for PHY-SI to simultaneously enhance the secrecy rate and the multicast rate. The UAV trajectory design in PHY-SI should be heavily dependent on several factors, e.g., the predetermined requirement on QoMS and the location of the ground nodes.

To characterize this new type of secrecy capacity region over a given flight duration, it is required to jointly optimize the UAV trajectory and power/user rate allocation over time subject to the UAV maximum speed and maximum transmit power constraints. The secrecy capacity region in this context has many different properties from its counterpart in the traditional terrestrial communications with fixed transmitter. In general, the two extreme points of the secrecy capacity region correspond to different UAV trajectories. Due to the maximum speed constraint, time sharing cannot be performed between these two extreme points for any finite UAV velocities. As a result, the secrecy capacity region is a nonconvex region, which is significantly different from the the one with a fixed transmitter. Moreover, the associated SCRM problem involves an infinite number of optimization variables, i.e., the time-varying transmit power and location of UAV, making this problem very challenging to solve.

\subsection{Other Potential Directions}
Compared to its counterpart in one-way networks, PHY-SI in two-way networks is less studied and hence may have greater research potential. In recent years, full duplex has received renewed interest with the advances of self-interference (SI) cancellation techniques (see \cite{liu2015band} for a detailed overview), and has been regarded as a promising physical-layer technology to meet the explosive data requirement for 5G. In bidirectional relay networks, full duplex has the potential to double the spectral efficiency by combining the MAC and BBC phases over the same RF bands. Concomitantly, it is necessary to explore the effects of SI in such a PHY-SI-enabled system. At the relay side, a common approach to eliminating SI is zero forcing (ZF) precoding, provided that the relay is equipped with multiple antennas and has perfect CSI of the SI link. At the side of the receiving nodes, if the nodes are equipped with a single antenna, the SI cannot be completely eliminated owing to issues such as the large dynamic range of the full-duplex SI\cite{liu2015band}, even though they have knowledge of its own transmitted signals. As such, the presence of residual SI will simultaneously undermine the performance of different services. Besides, it brings new challenges to the joint optimization of transmit beamformers to maximize the secrecy rate region.

On the other hand, as mentioned, the performance of PHY-SI is contingent on the available power in the network. Since PHY-SI intensifies the demand for energy, the issue of energy shortage is becoming increasingly prominent. Besides the design of energy-efficient transmission scheme, an alternative solution to this issue is simultaneous wireless information and power transfer (SWIPT), which explores a dual use of microwave signals to transfer information jointly with energy using the same waveform. To realize SWIPT in practice, the received signal has to be split in two distinct parts, one for energy harvesting and the other for information decoding. The signal splitting can be performed in different domains (time, power, antenna, and space), which gives rise to different techniques, i.e., time switching, power splitting, antenna switching and spatial switching\cite{krikidis2014simultaneous,bi2015wireless,chen2015enhancing,lu2015wireless}. In SWIPT-based PHY-SI system, one crucial problem lies in the increasing operational complexity. The receiver needs to perform successive information cancellation and energy harvesting simultaneously, which leads to large overhead. Consequently, the power splitting technique, which entails a high-complexity hardware implementation, might not be suitable for SWIPT-based PHY-SI. The other three techniques seem to be more suitable to SWIPT-based PHY-SI, as they decouple PHY-SI and energy harvesting and allow for a simpler operation at the receiver. Even so, the associated optimization problems are complicated and non-convexified by the integration of multiple service messages as compared to the case with single service, which is worth further studies on it.

\section{Conclusion}
This article introduced an emerging PHY-SI technique from both theoretical and technical perspectives. First, we introduced various PHY-SI schemes and revealed the central issues of PHY-SI in various scenarios, i.e., the characterization of tradeoff among different services. The general approaches to finding the capacity-achieving transmit covariance matrices were also mentioned. Then we put forth several state-of-the-art PHY-SI techniques. In particular, we introduced the AN-aided transmission and eigenmode transmission, both in the sense of PHY-SI. PHY-SI with robustness and energy efficiency was also discussed. It was shown that some interesting issues, along with several new insights, could emerge from these techniques. Finally, some potential research directions were identified for future work. It is hoped that this article could serve as a stepping-stone to the future efforts in these directions.

\bibliography{PHYSI_Survey}

\begin{thebibliography}{10}
\providecommand{\url}[1]{#1}
\csname url@samestyle\endcsname
\providecommand{\newblock}{\relax}
\providecommand{\bibinfo}[2]{#2}
\providecommand{\BIBentrySTDinterwordspacing}{\spaceskip=0pt\relax}
\providecommand{\BIBentryALTinterwordstretchfactor}{4}
\providecommand{\BIBentryALTinterwordspacing}{\spaceskip=\fontdimen2\font plus
\BIBentryALTinterwordstretchfactor\fontdimen3\font minus
  \fontdimen4\font\relax}
\providecommand{\BIBforeignlanguage}[2]{{%
\expandafter\ifx\csname l@#1\endcsname\relax
\typeout{** WARNING: IEEEtran.bst: No hyphenation pattern has been}%
\typeout{** loaded for the language `#1'. Using the pattern for}%
\typeout{** the default language instead.}%
\else
\language=\csname l@#1\endcsname
\fi
#2}}
\providecommand{\BIBdecl}{\relax}
\BIBdecl

\bibitem{lecompte2012evolved}
D.~Lecompte and F.~Gabin, ``Evolved multimedia broadcast/multicast service
  (e{MBMS}) in {LTE}-advanced: overview and {R}el-11 enhancements,''
  \emph{{IEEE} Commun. Mag.}, vol.~50, no.~11, pp. 68--74, Nov. 2012.

\bibitem{etemad2009multicast}
K.~Etemad and L.~Wang, ``Multicast and broadcast multimedia services in mobile
  {WiMAX} networks,'' \emph{{IEEE} Commun. Mag.}, vol.~47, no.~10, pp. 84--91,
  Oct. 2009.

\bibitem{andrew2014what}
J.~G. Andrews, S.~Buzzi, W.~Choi \emph{et~al.}, ``What will {5G} be?''
  \emph{{IEEE} J. Sel. Areas Commun.}, vol.~32, no.~6, pp. 1065--1082, Jun.
  2014.

\bibitem{costa1983writing}
M.~Costa, ``Writing on dirty paper,'' \emph{{IEEE} Trans. Inf. Theory},
  vol.~29, no.~3, pp. 439--441, May 1983.

\bibitem{spencer2004an}
Q.~H. Spencer, C.~B. Peel, A.~L. Swindlehurst, and M.~Haardt, ``An introduction
  to the multi-user {MIMO} downlink,'' \emph{{IEEE} Commun. Mag.}, vol.~42,
  no.~10, pp. 60--67, Oct. 2004.

\bibitem{sidi2006transmit}
N.~D. Sidiropoulos, T.~N. Davidson, and Z.-Q.~T. Luo, ``Transmit beamforming
  for physical-layer multicasting,'' \emph{{IEEE} Trans. Signal Process.},
  vol.~54, no.~6, pp. 2239--2251, Jun. 2006.

\bibitem{kim2011optimal}
H.~Kim, D.~J. Love, and S.~Y. Park, ``Optimal and successive approaches to
  signal design for multiple antenna physical layer multicasting,''
  \emph{{IEEE} Trans. Commun.}, vol.~59, no.~8, pp. 2316--2327, Aug. 2011.

\bibitem{wu2013physical}
S.~X. Wu, W.-K. Ma, and A.~M.-C. So, ``Physical-layer multicasting by
  stochastic transmit beamforming and {A}lamouti space-time coding,''
  \emph{{IEEE} Trans. Signal Process.}, vol.~61, no.~17, pp. 4230--4245, Sep.
  2013.

\bibitem{boneh2000why}
D.~Boneh, A.~Joux, and P.~Q. Nguyen, ``Why textbook {ElGamal} and {RSA}
  encryption are insecure,'' in \emph{Advances in Cryptology---ASIACRYPT},
  Kyoto, Japan, Dec. 2000, pp. 30--43.

\bibitem{wyner1975wire}
A.~D. Wyner, ``The wire-tap channel,'' \emph{The Bell System Technical J.},
  vol.~54, no.~8, pp. 1355--1387, Oct. 1975.

\bibitem{shiu2011physical}
Y.-S. Shiu, S.~Y. Chang, H.-C. Wu, S.~C.-H. Huang, and H.-H. Chen, ``Physical
  layer security in wireless networks: a tutorial,'' \emph{{IEEE} Wireless
  Commun.}, vol.~18, no.~2, pp. 66--74, Apr. 2011.

\bibitem{hong2013enhancing}
Y.-W.~P. Hong, P.-C. Lan, and C.-C.~J. Kuo, ``Enhancing physical-layer secrecy
  in multiantenna wireless systems: An overview of signal processing
  approaches,'' \emph{{IEEE} Signal Process. Mag.}, vol.~30, no.~5, pp. 29--40,
  Sep. 2013.

\bibitem{mukherjee2014principles}
A.~Mukherjee, S.~A. Fakoorian, J.~Huang, A.~L. Swindlehurst \emph{et~al.},
  ``Principles of physical layer security in multiuser wireless networks: A
  survey,'' \emph{{IEEE} Commun. Surveys Tuts.}, vol.~16, no.~3, pp.
  1550--1573, 2014.

\bibitem{liu2016physical}
Y.~Liu, H.-H. Chen, and L.~Wang, ``Physical layer security for next generation
  wireless networks: Theories, technologies, and challenges,'' \emph{{IEEE}
  Commun. Surveys Tuts.}, vol.~19, no.~1, pp. 347--376, 2017.

\bibitem{chen2017survey}
X.~Chen, D.~W.~K. Ng, W.~Gerstacker, and H.-H. Chen, ``A survey on
  multiple-antenna techniques for physical layer security,'' \emph{{IEEE}
  Commun. Surveys Tuts.}, vol.~19, no.~2, pp. 1027--1053, 2017.

\bibitem{Schaefer2014Physical}
R.~F. Schaefer and H.~Boche, ``Physical layer service integration in wireless
  networks: Signal processing challenges,'' \emph{{IEEE} Signal Process. Mag.},
  vol.~31, no.~3, pp. 147--156, Apr. 2014.

\bibitem{csiszar1978broadcast}
I.~Csisz{\'a}r and J.~K{\"o}rner, ``Broadcast channels with confidential
  messages,'' \emph{{IEEE} Trans. Inf. Theory}, vol.~24, no.~3, pp. 339--348,
  May 1978.

\bibitem{Hung2010Multiple}
H.~D. Ly, T.~Liu, and Y.~Liang, ``Multiple-input multiple-output {G}aussian
  broadcast channels with common and confidential messages,'' \emph{{IEEE}
  Trans. Inf. Theory}, vol.~56, no.~11, pp. 5477--5487, Oct. 2010.

\bibitem{khisti2010secure1}
A.~Khisti and G.~W. Wornell, ``Secure transmission with multiple antennas
  \uppercase\expandafter{\romannumeral1}: The {MISOME} wiretap channel,''
  \emph{{IEEE} Trans. Inf. Theory}, vol.~56, no.~7, pp. 3088--3104, Jul. 2010.

\bibitem{khisti2010secure2}
------, ``Secure transmission with multiple antennas
  \uppercase\expandafter{\romannumeral2}: The {MIMOME} wiretap channel,''
  \emph{{IEEE} Trans. Inf. Theory}, vol.~56, no.~11, pp. 5515--5532, Nov. 2010.

\bibitem{liu2009secrecy}
R.~Liu and H.~V. Poor, ``Secrecy capacity region of a multiple-antenna
  {G}aussian broadcast channel with confidential messages,'' \emph{{IEEE}
  Trans. Inf. Theory}, vol.~55, no.~3, pp. 1235--1249, Mar. 2009.

\bibitem{park2016weighted}
D.~Park, ``Weighted sum rate maximization of {MIMO} broadcast and interference
  channels with confidential messages,'' \emph{{IEEE} Trans. Wireless Commun.},
  vol.~15, no.~3, pp. 1742--1753, Mar. 2016.

\bibitem{fakoorian2013on}
S.~A.~A. Fakoorian and A.~L. Swindlehurst, ``On the optimality of linear
  precoding for secrecy in the {MIMO} broadcast channel,'' \emph{{IEEE} J. Sel.
  Areas Commun.}, vol.~31, no.~9, pp. 1701--1713, Sep. 2013.

\bibitem{geraci2012secrecy}
G.~Geraci, M.~Egan, J.~Yuan, A.~Razi, and I.~B. Collings, ``Secrecy sum-rates
  for multi-user {MIMO} regularized channel inversion precoding,'' \emph{{IEEE}
  Trans. Commun.}, vol.~60, no.~11, pp. 3472--3482, Nov. 2012.

\bibitem{ekrem2012capacity}
E.~Ekrem and S.~Ulukus, ``Capacity region of {G}aussian {MIMO} broadcast
  channels with common and confidential messages,'' \emph{{IEEE} Trans. Inf.
  Theory}, vol.~58, no.~9, pp. 5669--5680, Sep. 2012.

\bibitem{liu2013new}
R.~Liu, T.~Liu, H.~V. Poor, and S.~Shamai~(Shitz), ``New results on
  multiple-input multiple-output broadcast channels with confidential
  messages,'' \emph{{IEEE} Trans. Inf. Theory}, vol.~59, no.~3, pp. 1346--1359,
  Mar. 2013.

\bibitem{rodriguez2015physical}
L.~J. Rodriguez, N.~H. Tran, T.~Q. Duong, T.~Le-Ngoc, M.~Elkashlan, and
  S.~Shetty, ``Physical layer security in wireless cooperative relay networks:
  State of the art and beyond,'' \emph{IEEE Commun. Mag.}, vol.~53, no.~12, pp.
  32--39, Dec. 2015.

\bibitem{chen2015multi}
X.~Chen, C.~Zhong, C.~Yuen, and H.-H. Chen, ``Multi-antenna relay aided
  wireless physical layer security,'' \emph{IEEE Commun. Mag.}, vol.~53,
  no.~12, pp. 40--46, Dec. 2015.

\bibitem{he2010cooperation}
X.~He and A.~Yener, ``Cooperation with an untrusted relay: A secrecy
  perspective,'' \emph{{IEEE} Trans. Inf. Theory}, vol.~56, no.~8, pp.
  3807--3827, Aug. 2010.

\bibitem{jeong2012joint}
C.~Jeong, I.-M. Kim, and D.~I. Kim, ``Joint secure beamforming design at the
  source and the relay for an amplify-and-forward {MIMO} untrusted relay
  system,'' \emph{{IEEE} Trans. Signal Process.}, vol.~60, no.~1, pp. 310--325,
  Jan. 2012.

\bibitem{liu2015band}
G.~Liu, F.~R. Yu, H.~Ji, V.~C. Leung, and X.~Li, ``In-band full-duplex
  relaying: A survey, research issues and challenges,'' \emph{{IEEE} Commun.
  Surveys Tuts.}, vol.~17, no.~2, pp. 500--524, Second quarter 2015.

\bibitem{Wyrembelski2012Physical}
R.~Wyrembelski and H.~Boche, ``Physical layer integration of private, common,
  and confidential messages in bidirectional relay networks,'' \emph{{IEEE}
  Trans. Wireless Commun.}, vol.~11, no.~9, pp. 3170--3179, Sep. 2012.

\bibitem{mei2016secrecy}
W.~Mei, Z.~Chen, and J.~Fang, ``Secrecy capacity region maximization in
  {G}aussian {MISO} channels with integrated services,'' \emph{{IEEE} Signal
  Process. Lett.}, vol.~23, no.~8, pp. 1146--1150, Jul. 2016.

\bibitem{mei2017on}
W.~Mei, Z.~Chen, L.~Li, J.~Fang, and S.~Li, ``On artificial-noise-aided
  transmit design for multi-user {MISO} systems with integrated services,''
  \emph{{IEEE} Trans. Veh. Technol.}, vol.~66, no.~9, pp. 8179--8195, Sep.
  2017.

\bibitem{goel2008guaranteeing}
S.~Goel and R.~Negi, ``Guaranteeing secrecy using artificial noise,''
  \emph{{IEEE} Trans. Wireless Commun.}, vol.~7, no.~6, pp. 2180--2189, Jun.
  2008.

\bibitem{li2013spatially}
Q.~Li and W.-K. Ma, ``Spatially selective artificial-noise aided transmit
  optimization for {MISO} multi-{E}ves secrecy rate maximization,''
  \emph{{IEEE} Trans. Signal Process.}, vol.~61, no.~10, pp. 2704--2717, May
  2013.

\bibitem{li2013transmit}
Q.~Li, M.~Hong, H.-T. Wai, Y.-F. Liu, W.-K. Ma, and Z.-Q. Luo, ``Transmit
  solutions for {MIMO} wiretap channels using alternating optimization,''
  \emph{{IEEE} J. Sel. Areas Commun.}, vol.~31, no.~9, pp. 1714--1727, Sep.
  2013.

\bibitem{zheng2015multi}
T.-X. Zheng, H.-M. Wang, J.~Yuan, D.~Towsley, and M.~H. Lee, ``Multi-antenna
  transmission with artificial noise against randomly distributed
  eavesdroppers,'' \emph{{IEEE} Trans. Commun.}, vol.~63, no.~11, pp.
  4347--4362, Nov. 2015.

\bibitem{telatar1999capacity}
E.~Telatar, ``Capacity of multi-antenna {G}aussian channels,''
  \emph{Transactions on Emerging Telecommunications Technologies}, vol.~10,
  no.~6, pp. 585--595, 1999.

\bibitem{fakoorian2012optimal}
S.~A.~A. Fakoorian \emph{et~al.}, ``Optimal power allocation for {GSVD}-based
  beamforming in the {MIMO} gaussian wiretap channel,'' in \emph{Proc. {IEEE}
  Int. Symp. Inf. Theory}, Cambridge, MA, Jul. 2012, pp. 2321--2325.

\bibitem{mei2016GSVD}
W.~Mei, Z.~Chen, and J.~Fang, ``{GSVD}-based precoding in {MIMO} systems with
  integrated services,'' \emph{{IEEE} Signal Process. Lett.}, vol.~23, no.~11,
  pp. 1528--1532, Sep. 2016.

\bibitem{he2013wireless}
\BIBentryALTinterwordspacing
B.~He, X.~Zhou, and T.~D. Abhayapala, ``Wireless physical layer security with
  imperfect channel state information: A survey,'' Jun. 2013. [Online].
  Available: \url{http://arxiv.org/abs/1307.4146}
\BIBentrySTDinterwordspacing

\bibitem{wu2014green}
Y.~Wu, Y.~Chen, J.~Tang, D.~K. So \emph{et~al.}, ``Green transmission
  technologies for balancing the energy efficiency and spectrum efficiency
  trade-off,'' \emph{IEEE Commun. Mag.}, vol.~52, no.~11, pp. 112--120, Nov.
  2014.

\bibitem{wu2017overview}
Q.~Wu, G.~Y. Li, W.~Chen, D.~W.~K. Ng, and R.~Schober, ``An overview of
  sustainable green {5G} networks,'' \emph{IEEE Wireless Commun.}, vol.~24,
  no.~4, pp. 72--80, Aug. 2017.

\bibitem{zhang2017fundamental}
S.~Zhang, Q.~Wu, S.~Xu, and G.~Y. Li, ``Fundamental green tradeoffs:
  Progresses, challenges, and impacts on {5G} networks,'' \emph{{IEEE} Commun.
  Surveys Tuts.}, vol.~19, no.~1, pp. 33--56, 2017.

\bibitem{belmega2011energy}
E.~V. Belmega and S.~Lasaulce, ``Energy-efficient precoding for
  multiple-antenna terminals,'' \emph{{IEEE} Trans. Signal Process.}, vol.~59,
  no.~1, pp. 329--340, Jan. 2011.

\bibitem{xu2013energy}
J.~Xu and L.~Qiu, ``Energy efficiency optimization for {MIMO} broadcast
  channels,'' \emph{{IEEE} Trans. Wireless Commun.}, vol.~12, no.~2, pp.
  690--701, Feb. 2013.

\bibitem{zhang2014energy}
H.~Zhang, Y.~Huang, S.~Li, and L.~Yang, ``Energy-efficient precoder design for
  {MIMO} wiretap channels,'' \emph{{IEEE} Commun. Lett.}, vol.~18, no.~9, pp.
  1559--1562, Sep. 2014.

\bibitem{zappone2016energy}
A.~Zappone, P.-H. Lin, and E.~A. Jorswieck, ``Energy efficiency of confidential
  multi-antenna systems with artificial noise and statistical {CSI},''
  \emph{{IEEE} J. Sel. Topics Signal Process.}, vol.~10, no.~8, pp. 1462--1477,
  Dec. 2016.

\bibitem{ng2012energy}
D.~W.~K. Ng, E.~S. Lo, and R.~Schober, ``Energy-efficient resource allocation
  for secure {OFDMA} systems,'' \emph{{IEEE} Trans. Veh. Technol.}, vol.~61,
  no.~6, pp. 2572--2585, Jul. 2012.

\bibitem{chen2013energy}
X.~Chen and L.~Lei, ``Energy-efficient optimization for physical layer security
  in multi-antenna downlink networks with {QoS} guarantee,'' \emph{{IEEE}
  Commun. Lett.}, vol.~17, no.~4, pp. 637--640, Apr. 2013.

\bibitem{mei2017energy}
W.~Mei, Z.~Chen, and J.~Fang, ``Energy efficiency region for {G}aussian {MISO}
  channels with integrated services,'' \emph{{IEEE} Wireless Commun. Lett.},
  vol.~6, no.~1, pp. 90--93, Mar. 2017.

\bibitem{li2017constant}
Q.~Li, C.~Li, and J.~Lin, ``Constant modulus beamforming for large-scale
  {MISOME} wiretap channel,'' in \emph{Proc. 25th European Signal Processing
  Conference}, Kos, Greece, Aug. 2017, pp. 2541--2545.

\bibitem{qin2013power}
H.~Qin, Y.~Sun, T.-H. Chang, X.~Chen, C.-Y. Chi, M.~Zhao, and J.~Wang, ``Power
  allocation and time-domain artificial noise design for wiretap {OFDM} with
  discrete inputs,'' \emph{{IEEE} Trans. Wireless Commun.}, vol.~12, no.~6, pp.
  2717--2729, Jun. 2013.

\bibitem{benfarah2016power}
A.~Benfarah, S.~Tomasin, and N.~Laurenti, ``Power allocation in multiuser
  parallel {G}aussian broadcast channels with common and confidential
  messages,'' \emph{{IEEE} Trans. Commun.}, vol.~64, no.~6, pp. 2326--2339,
  Jun. 2016.

\bibitem{krikidis2014simultaneous}
I.~Krikidis, S.~Timotheou, S.~Nikolaou, G.~Zheng, D.~W.~K. Ng, and R.~Schober,
  ``Simultaneous wireless information and power transfer in modern
  communication systems,'' \emph{IEEE Commun. Mag.}, vol.~52, no.~11, pp.
  104--110, Nov. 2014.

\bibitem{bi2015wireless}
S.~Bi, C.~K. Ho, and R.~Zhang, ``Wireless powered communication: Opportunities
  and challenges,'' \emph{IEEE Commun. Mag.}, vol.~53, no.~4, pp. 117--125,
  Apr. 2015.

\bibitem{chen2015enhancing}
X.~Chen, Z.~Zhang, H.-H. Chen, and H.~Zhang, ``Enhancing wireless information
  and power transfer by exploiting multi-antenna techniques,'' \emph{IEEE
  Commun. Mag.}, vol.~53, no.~4, pp. 133--141, Apr. 2015.

\bibitem{lu2015wireless}
X.~Lu, P.~Wang, D.~Niyato, D.~I. Kim, and Z.~Han, ``Wireless networks with {RF}
  energy harvesting: A contemporary survey,'' \emph{{IEEE} Commun. Surveys
  Tuts.}, vol.~17, no.~2, pp. 757--789, Second quarter 2015.

\end{thebibliography}
\bibliographystyle{IEEEtran}
\end{document}